\documentclass[a4paper,12pt]{article}  
\usepackage{amsfonts}
\usepackage{amsmath}
\usepackage{amssymb}
\usepackage{cite}
\usepackage{appendix}
\usepackage{array}
\usepackage{cases}
\usepackage{enumitem}
\usepackage{graphicx}
\usepackage{hyperref}
\usepackage{mathrsfs}
\hypersetup{colorlinks, linkcolor={red},citecolor={blue},urlcolor={blue}} 
\numberwithin{equation}{section}

\newcommand{\qm}[1]{``#1''}
\def\d{{\rm d}} 
\def\nn{\nonumber} 

\title{Discontinuous normals in non-Euclidean geometries and two-dimensional gravity}

\date{\today}

\author{
Emmanuele Battista ORCID: 0000-0001-5361-7109 \\
Department of Physics, University of Vienna, \\
Boltzmanngasse 5, A-1090 Vienna, Austria\\
Giampiero Esposito ORCID: 0000-0001-5930-8366 \\
Dipartimento di Fisica ``Ettore Pancini''
and \\
INFN Sezione di Napoli,
Complesso Universitario di Monte S. Angelo, \\
Via Cintia Edificio 6, 80126 Napoli, Italy}

\begin{document}

\maketitle

\begin{abstract}
This paper builds two detailed examples of generalized normal in 
non-Euclidean spaces, i.e. the hyperbolic and elliptic geometries. In the
hyperbolic plane we define a $n$-sided hyperbolic polygon
$\mathscr{P}$, which is the Euclidean closure of the hyperbolic
plane $\mathbb{H}$, bounded by $n$ hyperbolic geodesic segments. The 
polygon $\mathscr{P}$ is built by considering the unique geodesic that
connects the $n+2$ vertices ${\tilde z},z_{0},z_{1},...,z_{n-1},z_{n}$.
The geodesics that link the vertices are Euclidean semicircles
centred on the real axis. The vector normal to the geodesic linking
two consecutive vertices is evaluated and turns out to be discontinuous. 
Within the framework of elliptic geometry, we solve the geodesic equation 
and construct a geodesic triangle. Also in this case, we obtain a 
discontinuous normal vector field. Last, the possible application to
two-dimensional Euclidean quantum gravity is outlined. 
\end{abstract}

\section{Introduction}
\setcounter{equation}{0}
\label{Sec:Intro}

Suppose we want to measure the perimeter of a plane curve and we
only have at our disposal a ruler, but not a whire that would match
perfectly the boundary \cite{DeGiorgi}. 
How are we going to evaluate the perimeter?
One can indeed look for polygons whose sides are measurable with
the help of the ruler, and one can ``approximate'' the shape of
the curve by means of such polygons, asking ourselves the question
by how much such polygons are close to the figure whose perimeter 
we want to evaluate. Within this framework, the very fruitful idea
of Caccioppoli \cite{RC1,RC2} 
was to ``measure'' the distance between the polygon 
and the figure by means of the area of the difference set. In order
to gain an idea of the perimeter, we have to take a sequence of polygons,
such that the area of the difference set becomes smaller, and they
will provide a sequence of approximate values of the perimeter. 
However, the area-type approximation of the figure can be obtained
also by means of polygons having a perimeter unnecessarily large
(for example, if we curl a polygon, we can achieve a very large
perimeter). Thus, among all sequences of polygons which approximate
a geometric figure, we must take, among the limits of their
perimeters, the smallest value, i.e., the minimal limit of
perimeters of the approximating polygons.

The development of geometric measure theory 
\cite{RC1,RC2,DG1,DG2,book1,book2,book3,book4,book5,book6,book7}
led therefore to the discovery of many important concepts including,
in particular, finite-perimeter sets with their reduced boundary
(see definitions in  \ref{Appendix:Finite-Perimeter-Set}), two concepts
that play an important role in modern mathematics. With hindsight, 
one can say that the divergence theorem does not hold on the 
topological boundary of a finite-perimeter set, but only on the
reduced boundary, which is therefore the truly important concept
of boundary. Let us here review some key aspects of this framework, 
which provide a motivation of our research.

If $E \subset {\mathbb{R}}^{n}$ is a finite-perimeter set, 
the De Giorgi structure theorem ensures that its reduced
boundary $\partial^{*}E$ can be written in the form
\begin{equation}
\partial^{*}E=\cup_{l=1}^{\infty}K_{l} \cup N_{0},
\label{(1.1)}
\end{equation}
where the $K_{l}$ are disjoint compact sets, and $N_{0}$ is
a set having vanishing $(n-1)$-dimensional Hausdorff measure.
The sets $K_{l}$ are contained in $(n-1)$-dimensional 
manifolds $M_{l}$ of class $C^{1}$. If the point $x$ belongs
to $K_{l}$, one finds that the generalized normal to $E$,
denoted by $\nu_{E}$, can be obtained by the equation
\begin{equation}
\nu_{E}(x)=\nu_{M_{l}}(x).
\label{(1.2)}
\end{equation}
This means that the normal $\nu_{E}$ at $x$ is given by
the normal existing at $x$ on the manifold $M_{l}$ (for this
purpose, it is of crucial importance that the manifold $M_{l}$
should be of class $C^{1}$). Thus, the generalized normal is
evaluated from the usual normal to a countable infinity of
compact portions $K_{l}$ of smooth manifolds $M_{l}$. 

However, despite this profound result, the actual evaluation 
of the generalized normal may turn out to be impossible in some
cases. In order to understand this feature, one can consider
a dense sequence $\left \{ q_{k} \right \}$ of points of ${\mathbb{R}}^{n}$.
If the points $q_{k}$ are vectors with $n$ components 
which have rational coordinates
$$
\left({q_{k}}_{1},{q_{k}}_{2},...,{q_{k}}_{n}\right),
$$
then within each ball $B_{1 \over k}(q_{k})$ centred at the
point $q_{k} \in {\mathbb{R}}^{n}$ and having radius 
${1 \over k}$ one can find infinitely many points, and the same
is true outside such a ball. The desired set $E$ is in this case
\begin{equation}
E=\cup_{k=1}^{\infty} B_{1 \over k}(q_{k}),
\label{(1.3)}
\end{equation}
and its volume can be majorized as follows:
\begin{equation}
{\rm Vol}(E) \leq \sum_{k=1}^{\infty} {\rm Vol}
\left(B_{1 \over k}(q_{k}) \right)
={\rm Vol}(B_{1}) \sum_{k=1}^{\infty}
{1 \over k^{n}} < \infty,
\label{(1.4)}
\end{equation}
where $B_{1}$ is the ball of unit radius. We have therefore defined
a finite-volume set. Although no picture can be drawn, one can think
of $E$ as consisting of infinitely many balls about the points 
$q_{k}$ of ${\mathbb{R}}^{n}$, which are dense in ${\mathbb{R}}^{n}$.

The topological boundary $\partial E$ of $E$ coincides with the
whole of ${\mathbb{R}}^{n}$ minus $E$. In fact, given a point
$x$ of ${\mathbb{R}}^{n}$, one can find a sequence of points
with rational coordinates, extracted from the sequence 
$\left \{ q_{k} \right \}$ considered before. Such an extracted 
sequence tends to $x \in {\mathbb{R}}^{n}$ because the points 
with rational coordinates are dense. This means that 
$x \in {\mathbb{R}}^{n}$ is an accumulation point for $E$. Thus
$x$ belongs to the closure ${\overline E}$ of $E$. In turn,
since $E$ is an open set (being formed by a countable union
of open balls), the closure of $E$ is given by
\begin{equation}
{\overline E}=E \cup {\partial E},
\label{(1.5)}
\end{equation}
and hence
\begin{equation}
{\partial E}={\mathbb{R}}^{n}-E
\Longrightarrow 
{\rm Vol}({\partial E})=+ \infty.
\label{(1.6)}
\end{equation}

Let us now prove that $E$ is a finite-perimeter set, despite the
fact that its topological boundary has infinite volume. Indeed,
by definition the perimeter $P(E)$ is given by
\begin{equation}
P(E) \equiv {\rm sup} \left \{ \int_{E}{\rm div}T \; {\rm d}x:
\; \left \| T \right \|_{L^{\infty}} \leq 1, \;
T \in C_{c}^{1}({\mathbb{R}}^{n},{\mathbb{R}}^{n}) \right \},
\label{(1.7)}
\end{equation}
where the integral over $E$ is a volume integral, and the
defining conditions mean that $T$ is a vector field of 
$L^{\infty}$ norm never bigger than $1$, of class $C^{1}$
on ${\mathbb{R}}^{n}$ and having compact support on 
${\mathbb{R}}^{n}$. By virtue of the definition \eqref{(1.3)}, one
can write the volume integral of the divergence of $T$ in
the form
\begin{equation}
\int_{E}{\rm div} T \; {\rm d}x = 
\lim_{m \to \infty}
\int_{\cup_{k=1}^{m}B_{1 \over k}(q_{k})}
{\rm div}T \; {\rm d}x.
\label{(1.8)}
\end{equation}
The union of a finite number of balls 
$B_{1 \over k}(q_{k})$ gives rise to an irregular set, because
the balls do not have smooth intersections. Nevertheless, 
on such a set the divergence theorem holds, since such a set
is piecewise smooth. One can therefore write, by virtue of the
divergence theorem:
\begin{equation}
\lim_{m \to \infty}
\int_{\cup_{k=1}^{m}B_{1 \over k}(q_{k})}
{\rm div}T \; {\rm d}x = \lim_{m \to \infty}
\int_{\partial \left(\cup_{k=1}^{m} 
B_{1 \over k}(q_{k}) \right)}
T \cdot \nu \; {\rm d}H^{n-1},
\label{(1.9)}
\end{equation}
where $\nu$ is the normal to the topological boundary of
the union of open balls, and $dH^{n-1}$ is a surface measure.
By virtue of \eqref{(1.8)} and \eqref{(1.9)}, one finds the majorization
($H^{n-1}$ being the measure of the surface given by the
topological boundary of the union of open balls 
$B_{1 \over k}(q_{k})$)
\begin{eqnarray}
\; & \; & 
\int_{E}{\rm div}T \; {\rm d}x \leq \lim_{m \to \infty}
H^{n-1}\left(\partial \left( \cup_{k=1}^{m}
B_{1 \over k}(q_{k})\right) \right)
\nonumber \\
& \leq & \lim_{m \to \infty} \sum_{k=1}^{m}
H^{n-1} \left(\partial B_{1 \over k}(q_{k})\right),
\label{(1.10)}
\end{eqnarray}
where, on the first line, we have exploited the majorization
of the $L^{\infty}$ norm of $T$ and the unit norm of $\nu$,
while on the second line we have exploited the property
\begin{equation}
\partial \left(\cup_{k=1}^{m}B_{1 \over k}(q_{k})\right)
\leq \cup_{k=1}^{m} \left(\partial B_{1 \over k}(q_{k})\right).
\label{(1.11)}
\end{equation}
This latter condition means that the topological boundary of the
union of balls $B_{1 \over k}(q_{k})$ is in general smaller than
the union of the various topological boundaries of such balls.
Furthermore, we can re-express the majorization \eqref{(1.10)} as follows:
\begin{eqnarray}
\; & \; & \int_{E} {\rm div}T \; {\rm d}x \leq \lim_{m \to \infty}
H^{n-1}\left (\partial B_{1} \right)
\sum_{k=1}^{m} {1 \over k^{n-1}}  \leq  M, \quad {\rm if} \; n \geq 3.
\label{(1.12)}
\end{eqnarray}
Indeed, if $n \geq 3$, one can find a constant $M$ for which
$$
\sum_{k=1}^{m}{1 \over k^{n-1}} \leq M.
$$
By virtue of \eqref{(1.12)}, a real number $M$ exists such that
\begin{equation}
\int_{E}{\rm div}T \; {\rm d}x \leq M \Longrightarrow P(E) \leq M,
\label{(1.13)}
\end{equation}
i.e. the set $E$ has a finite perimeter.

To sum up, the set $E$ defined in Eq. \eqref{(1.3)} is a finite-perimeter set
of finite volume, while its topological boundary has infinite volume. 
Moreover, another consequence of De Giorgi's structure theorem is 
that, if $E$ is a finite-perimeter set and if $T$ is a vector field
of class $C^{1}$, then
\begin{equation}
\int_{E}{\rm div}T=\int_{\partial^{*}E}T \cdot \nu_{E}
\; {\rm d}H^{n-1}.
\label{(1.14)}
\end{equation}
Thus, the divergence theorem holds for a finite-perimeter set, but 
it does not involve the whole of the topological boundary
$\partial E$ (because $\partial E$ might contain parts that are
too irregular to see the divergence of $T$). The divergence theorem
involves therefore only the reduced boundary $\partial^{*}E$, which
represents the truly useful part of the topological boundary 
$\partial E$. With hindsight, this is the ultimate meaning of 
De Giorgi's structure theorem and of the concept of reduced boundary.
The finite-perimeter sets are therefore the most general objects 
for which the divergence theorem still holds.

In the example we have investigated it is impossible
to evaluate explicitly the generalized normal. Nevertheless, 
by virtue of De Giorgi's structure theorem, we know some qualitative
properties of great value: the reduced boundary is given by the
countable union of pieces of manifolds of class $C^{1}$, and the
generalized normal is given by the normal to such manifolds. In
such an example the reduced boundary is given by the countable
union of pieces $K_l$ of smooth manifolds, and such pieces are 
represented by awkward-looking pieces of spherical surfaces.  
One can also prove, for this example, the inclusion property
\begin{equation}
\partial^{*}E \subset \cup_{k=1}^{\infty} \partial
B_{1 \over k}(q_{k}),
\label{(1.15)}
\end{equation}
i.e., the reduced boundary is a subset of the countable union of the
topological boundaries of all open balls centred at $q_{k}$ and
having radius ${1 \over k}$. 

It is therefore clear that every explicit construction of generalized
normal is a challenging task which is nevertheless necessary  
if one wants to go beyond the purely qualitative properties.
The following sections are devoted to two original examples, motivated
by the desire to study geometric measure theory in non-Euclidean spaces. 
In particular, in Sec. \ref{Sec:hyperbolic-plane} we deal with hyperbolic 
geometry, whereas the case of elliptic geometry is investigated in Sec. 
\ref{Sec:elliptic-metric}. The physical relevance of our framework is
discussed in Sec. \ref{Sec:Application}, 
where we consider possible applications to the
action principle for two-dimensional Euclidean quantum gravity. In
particular, we deal with the Euclidean two-dimensional 
Callan-Giddings-Harvey-Strominger (hereafter referred to as CGHS) dilaton 
gravity model \cite{CGHS1992,Harvey1992,Strominger1995} (see also 
Refs. \cite{alpha,beta,gamma,delta}). As we will show, a two-dimensional
model makes it possible to overcome the issue regarding the definition of
the normal vector in higher-dimensional theories such as general relativity
(for recent reviews of general relativity and beyond, see, e.g., 
Refs. \cite{a,b,c} and references therein). Last, concluding remarks 
are made in Sec. \ref{Sec:Conclusions}.

\section{An example of generalized normal in the hyperbolic plane} 
\setcounter{equation}{0}
\label{Sec:hyperbolic-plane}

The hyperbolic (or Lobachevsky) plane \cite{Katok1992,Anderson2005,Needham1999} is defined as 
the upper half-plane $\mathbb{H}$ in the complex plane $\mathbb{C}$
\begin{equation}\label{upper-half-plane}
\mathbb{H} = \{ z \in \mathbb{C} \; : \; {\rm Im}(z) >0\},
\end{equation}
endowed with the metric
\begin{equation}\label{hyperbolic-metric}
g_{ab} {\rm d}x^a \otimes {\rm d}x^b= \dfrac{{\rm d}x \otimes {\rm d}x 
+{\rm d}y \otimes {\rm d}y }{y^2}, \qquad (a,b=1,2).
\end{equation}

Geodesics (or hyperbolic lines) in $\mathbb{H}$ are defined in terms of Euclidean objects 
in $\mathbb{C}$, being represented either by (the intersection of $\mathbb{H}$ with) 
Euclidean segments in $\mathbb{C}$ perpendicular to the real axis $\mathbb{R} 
= \{ z \in \mathbb{C} \; : \; {\rm Im}(z)=0 \}$ or by (the intersection of $\mathbb{H}$ with)  
Euclidean circles\footnote{Strictly speaking, we should distinguish 
between a circle and the associated circumference.} 
in $\mathbb{C}$  having Euclidean center on $\mathbb{R}$ and Euclidean 
radius $r$. Any two points in $\mathbb{H}$ can be joined by a unique geodesic.  
Let $\varphi \; : \; I \subset \mathbb{R} \rightarrow \mathbb{H}$ be the piecewise differentiable path
\begin{equation} \label{hyperbolic-geodesic}
\varphi = \{ z (t) = x(t) + {\rm i} y(t) \in \mathbb{H} \; : \; t \in I \},
\end{equation}
defining a geodesic in $\mathbb{H}$. Bearing in mind (\ref{hyperbolic-metric}), one finds that 
the geodesic (\ref{hyperbolic-geodesic}) satisfies the following 
system of ordinary differential equations \cite{Nakahara}:
\begin{subnumcases}{\label{hyperbolic-geodesic-eqs}}
y \ddot{x}  - 2 \dot{x} \dot{y}=0, & \label{hyperbolic-geodesic-eqs_a}\\ 
y \ddot{y} + \left(\dot{x}\right)^2 - \left(\dot{y}\right)^2=0, & \label{hyperbolic-geodesic-eqs_b}\
\end{subnumcases}
where $\dot{x}(t) \equiv \dfrac{{\rm d}x(t)}{{\rm d}t}$ and $\dot{y}(t) \equiv 
\dfrac{{\rm d}y(t)}{{\rm d}t}$. When $\dot{x}(t) =0 \; \forall\, t \in I$, the solution 
of (\ref{hyperbolic-geodesic-eqs}) is given by
\begin{equation}
z(t) = \left( a,b \,{\rm e}^{\pm t} \right), \qquad a \in \mathbb{R}, \; b >0,
\end{equation}
representing a vertical line in $\mathbb{H}$. On the other hand, if $\dot{x}(t)$ is nonvanishing, 
the system (\ref{hyperbolic-geodesic-eqs}) leads to 
\begin{equation}
z(t) = \left(r \tanh t +c, \dfrac{r}{\cosh t} \right), \qquad r>0, \; c \in \mathbb{R},
\end{equation}
which describes a Euclidean positive semicircle with Euclidean center at $(c,0)$ and Euclidean 
radius $r$, since $z(t)=(x(t),y(t))$ satisfies
\begin{equation}
\left(x(t)-c\right)^2 + y^2(t) = r^2.
\end{equation} 
It is not difficult  to show that the Euclidean center $(c,0)$ and the Euclidean radius $r$
of the Euclidean circle through $p,q \in \mathbb{H}$ can be written as   \cite{Anderson2005}
\begin{subequations}
\label{Euclidean-circle-c-and-r}
\begin{align}
c & = \dfrac{1}{2} \left( \dfrac{\vert p \vert^2 - \vert q \vert^2 }{{\rm Re}(p)-{\rm Re}(q)}\right), 
\label{Euclidean-circle-c} 
\\
r &= \vert c - p\vert = \vert c - q\vert.
\label{Euclidean-circle-r}
\end{align}
\end{subequations}

The hyperbolic length $\ell_{\mathbb{H}} (\gamma)$ of a piecewise differentiable path 
\begin{equation} 
\gamma= \{ z (t) = x(t) + {\rm i} y(t) \in \mathbb{H} \; : \; t \in I \subset \mathbb{R} \},
\end{equation}
is given by
\begin{equation}
\ell_{\mathbb{H}} (\gamma)= \int_{I} \dfrac{{\rm d}t}{y(t)}  
\left\vert \dfrac{{\rm d}z(t)}{{\rm dt}} \right\vert.
\end{equation}
The hyperbolic distance   $\rho_{\mathbb{H}} (z,w)$ between two points 
$z,w \in \mathbb{H}$ is defined by the formula
\begin{equation} \label{hyperbolic-distance}
\rho_{\mathbb{H}} (z,w) = {\rm inf } \; \ell_{\mathbb{H}} (\gamma),
\end{equation}
where the infimum is taken over all paths $\gamma$ joining $z,w \in \mathbb{H}$. 
Equation (\ref{hyperbolic-distance}) defines a distance function on $\mathbb{H}$, since  
it is non-negative, symmetric, and satisfies the triangle inequality \cite{Katok1992}. 
Moreover, it can be shown that for any $z,w \in \mathbb{H}$ \cite{Katok1992}\footnote{When 
the points $z,w \in \mathbb{H}$ are such that ${\rm Re} \left(z\right)= {\rm Re} \left(w\right)$, 
their hyperbolic distance reads as \cite{Katok1992,Anderson2005}
\begin{equation*} 
\left. \rho_{\mathbb{H}} (z,w) \right \vert_{{\rm Re} \left(z\right)= {\rm Re} \left(w\right)} 
= \log \left( \dfrac{{\rm Im}\left(w\right)}{{\rm Im}\left(z\right)}\right), 
\qquad \left({\rm Im}\left(w\right) > {\rm Im}\left(z\right) \right).
\end{equation*}}
\begin{equation} \label{hyperbolic_distance-1}
\rho_{\mathbb{H}} (z,w) = \log \left( \dfrac{\vert z - \bar{w} \vert + \vert z - w \vert }{\vert z 
- \bar{w} \vert - \vert z - w \vert }\right).
\end{equation}
The hyperbolic distance $\rho_{\mathbb{H}} (z,w) $ can also be written in terms of the 
Euclidean radius $r$ and the Euclidean center $(c,0)$ of the geodesic 
connecting $z,w \in \mathbb{H}$ as \cite{Anderson2005}
\begin{equation} \label{hyperbolic_distance-2}
\rho_{\mathbb{H}} (z,w) = \left \vert  \log \left[ \dfrac{\bigl(w-(c+r)
\bigr)\bigl(z-(c-r)\bigr)}{\bigl(w-(c-r)\bigr)\left(z-(c+r)\right)}\right]  \right \vert. 
\end{equation}

The group of all isometries of $\mathbb{H}$ is isomorphic to 
the space ${\rm PS} {\rm L}(2,\mathbb{R})$,
which is defined as follows:
$$
{\rm PSL}(2,\mathbb{R}) \equiv \left \{
z \rightarrow T(z)={(az+b)\over (cz+d)} | (ad-bc=1) \right \}.
$$
This is equivalent to expressing ${\rm PSL}(2,\mathbb{R})$ as the quotient
space ${\rm SL}(2,\mathbb{R})/Z_{2}$ because, if we change sign to all matrix
entries, both $T(z)$ and the determinant condition are preserved. Besides being
a group, ${\rm PSL}(2,\mathbb{R})$ is also a topological space in which the
fractional linear transformation $T(z)$ can be identified with the point
$(a,b,c,d)$ of ${\mathbb{R}}^{4}$. More precisely, as a topological space,
${\rm SL}(2,\mathbb{R})$ can be identified with the following subset of
${\mathbb{R}}^{4}$:
$$
X \equiv \left \{ (a,b,c,d) \in {\mathbb{R}}^{4}: ad-bc=1 \right \}.
$$
If one defines
$$
\delta(a,b,c,d) \equiv (-a,-b,-c,-d),
$$
the map $\delta: X \rightarrow X$ is therefore a homeomorphism, and we can
write that
$$
{\rm PSL}(2,\mathbb{R}) = {\rm SL}(2,\mathbb{R})/ \delta,
$$
which is a more precise expression of the quotient space formula involving 
$Z_{2}$. Therefore ${\rm PSL}(2,\mathbb{R})$ is a topological group,
and the fractional linear maps $T$ have a norm induced from 
${\mathbb{R}}^{4}$ given by
$$
\left \| T \right \| \equiv \sqrt{a^{2}+b^{2}+c^{2}+d^{2}}.
$$
A hyperbolic $n$-sided  polygon $\mathscr{P}$ is a closed set of $\mathbb{H} \cup 
\mathbb{R} \cup \{ \infty \}$ (i.e., the Euclidean closure of $\mathbb{H}$) bounded 
by $n$ hyperbolic geodesic segments. The vertices of $\mathscr{P}$, defined as the 
points of intersection of two line segments, can lie in $ \mathbb{R} \cup \{ \infty \}$, 
although no segment of $\mathbb{R}$ can belong to $\mathscr{P}$ \cite{Katok1992}. 

An example which makes it possible for us to display an explicit expression for 
the dual normal can be constructed as follows. Let
\begin{subequations}
\label{points-hyperbolic-polygon}
\begin{align}
\tilde{z} &= \left(0,1\right),
\\
z_0 &= \left(1,1\right),
 \\
z_1 &= \left(1+\dfrac{1}{3},\dfrac{1}{2}\right),
 \\
z_2 &= \left(1+\dfrac{1}{3} + \dfrac{1}{9},\dfrac{1}{4}\right),
 \\
& \vdots
\nonumber \\
z_{n-1} &= \left(1+\dfrac{1}{3} + \dfrac{1}{9} + \dots + \dfrac{1}{3^{n-1}},
\dfrac{1}{2^{n-1}}\right), \label{point-z-n-1}
 \\
z_{n} &= \left(1+\dfrac{1}{3} + \dfrac{1}{9} + \dots + \dfrac{1}{3^{n-1}} 
+ \dfrac{1}{3^n},\dfrac{1}{2^{n}}\right)\label{point-z-n}
\end{align}
\end{subequations}
denote the $n+2$ vertices of a hyperbolic $(n+2)$-sided  polygon $\mathscr{P}$. Such a polygon 
can be constructed by considering the unique geodesic connecting the following 
pairs of points:  $\tilde{z}$ and $z_0$, $z_0$ and $z_1$, $z_1$ and $z_2$, ..., $z_{n-1}$ 
and $z_n$, and eventually  $z_n$ and $\tilde{z}$. Since the points occurring in Eq. 
(\ref{points-hyperbolic-polygon}) have different real parts, all the aforementioned 
geodesics will be Euclidean positive semicircles having center on the real axis $\mathbb{R}$.  
The generic geodesic $\varphi_n$ joining $z_{n-1}$ and $z_n$ will be defined by 
\begin{equation}\label{geodesic-varphi_n}
\varphi_n = \{ z \in \mathbb{H} \; : \;   f(z)\equiv \vert z- c_n \vert^2 -\left( r_n\right)^2=0 \},
\end{equation}
where,  from Eq. (\ref{hyperbolic-geodesic-eqs}), we have
\begin{align}
c_n &= \dfrac{1}{2}  \left(\dfrac{\vert z_{n-1}\vert^2 - \vert z_{n}\vert^2}{{\rm Re}
\left(z_{n-1}\right)-{\rm Re}\left(z_{n}\right)}\right),
\\
r_n&= \vert c_n- z_{n-1} \vert =\vert c_n- z_{n} \vert.
\end{align}
Therefore, we find that 
\begin{align}
c_n &= \dfrac{1}{3^n} \sum_{j=1}^{n} 3^j + \alpha_n,
\\
r_n &=  \sqrt{\left( \alpha_n \right)^2 +\dfrac{1}{2^{2n-2}}},
\end{align}
where
\begin{equation}
\alpha_n \equiv \dfrac{2^{2n}-3^{2n+1}}{3^n \, 2^{2n+1}}.
\end{equation}
The unit normal vector $N_{(n)}$  to the geodesic (\ref{geodesic-varphi_n}) will be
\begin{equation} \label{normal-N}
N_{(n)}= \left(N_{(n)}^1, N_{(n)}^2 \right)= \dfrac{y}{r_n} 
\left(x-c_n,y \right), \; \; \qquad (n \in \mathbb{N}),
\end{equation}
with $x,y$ subjected to the conditions $\left(x-c_n\right)^2 + y^2 = \left(r_n\right)^2$ and $y>0$. 

We note that
\begin{align} \label{limit-n-to-infinity-1}
&\lim_{n \to \infty} c_n = \dfrac{3}{2},
\nonumber \\
&\lim_{n \to \infty} r_n =0,
\end{align}
which define an Euclidean degenerate circle. From the above equation jointly with 
formula (\ref{hyperbolic_distance-2}), we find that 
\begin{equation} \label{distance-z-n-1-and-z-n-large-n}
\rho_{\mathbb{H}} \left(z_{n-1},z_{n}\right) \xrightarrow[n \to \infty]{} 0.
\end{equation}
We note that the above result cannot be obtained by employing Eq. (\ref{hyperbolic_distance-1}), since
\begin{equation} \label{z-n-with-large-n}
z_n \xrightarrow[n \to \infty]{}  \left(\dfrac{3}{2},0 \right),  
\end{equation}
meaning that $z_n$  does not belong to $\mathbb{H}$ when $n \to \infty$ (see  
\ref{Appendix-log-2} for further details). In this limit,  Eq. (\ref{hyperbolic_distance-1}) 
would lead to a meaningless result.

The geodesic $\tilde{\varphi}$ connecting $\tilde{z}$ and $z_0$ 
\begin{equation} \label{geodesic-tilde-phi}
\tilde{\varphi} = \{ z \in \mathbb{H} \; : \;   f(z)\equiv \vert z- \tilde{c} \vert^2 - \tilde{r}^2=0 \},
\end{equation}
will have Euclidean center at $(\tilde{c},0)$ and Euclidean radius $\tilde{r}$, whereas for the 
geodesic $\hat{\varphi}$ through $z_n$ and $\tilde{z}$ 
\begin{equation} \label{geodesic-hat-phi}
\hat{\varphi} = \{ z \in \mathbb{H} \; : \;   f(z)\equiv \vert z- \hat{c} \vert^2 - \hat{r}^2=0 \},
\end{equation}
the  Euclidean center lies at $(\hat{c},0)$ and the Euclidean radius is $\hat{r}$. From  Eq. 
(\ref{Euclidean-circle-c-and-r}), we find
\begin{align}
\tilde{c} & = \dfrac{1}{2}, 
\nonumber \\
\tilde{r} &= \dfrac{\sqrt{5}}{2},
\end{align}
and
\begin{align}
\hat{c} &= \dfrac{3}{4} \left(1-\dfrac{1}{3^{n+1}}\right) +\dfrac{3^n 
\left(1-2^{2n}\right)}{2^{2n} \left(3^{n+1}-1\right)},
\nonumber \\
\hat{r} &= \sqrt{\hat{c}^2 +1}.
\end{align}
In the limit of large $n$, we obtain 
\begin{align}
& \lim_{n \to \infty} \hat{c} = \dfrac{5}{12}, 
\nonumber \\
& \lim_{n \to \infty} \hat{r} = \dfrac{13}{12},
\end{align}
which agree with Eq. (\ref{z-n-with-large-n}) (cf. also Eq. (\ref{limit-n-to-infinity-1})). 

The unit normal vector to the geodesic (\ref{geodesic-tilde-phi}) will be
\begin{equation} \label{normal-tilde-N}
\tilde{N} = \left(\tilde{N}^1, \tilde{N}^2\right)= \dfrac{y}{\tilde{r}} \left(x-\tilde{c},y\right),
\end{equation}
$x,y$ satisfying the conditions $\left(x-\tilde{c}\right)^2 + y^2 = \left(\tilde{r}\right)^2$ 
and $y>0$, whereas the unit normal vector to (\ref{geodesic-hat-phi}) is
\begin{equation} \label{normal-hat-N}
\hat{N} = \left(\hat{N} ^1, \hat{N} ^2\right)= \dfrac{y}{\hat{r}} \left(x-\hat{c},y\right),
\end{equation}
where $x,y$ are such that the relations $\left(x-\hat{c}\right)^2 + y^2 = \left(\hat{r}\right)^2$ 
and $y>0$ are fulfilled. 

The polygon $\mathscr{P}$ having vertices (\ref{points-hyperbolic-polygon}) has thus a generalized 
normal represented by Eqs. (\ref{normal-N}), (\ref{normal-tilde-N}), and (\ref{normal-hat-N}), i.e.,
\begin{equation}
\nu_E = \left\{ N_{(n)}, \tilde{N}, \hat{N} \right\}, \; \; \qquad (n \in \mathbb{N}).
\label{generalized-normal-hyperbolic}
\end{equation}
It follows from the above equation that the $n$-th segment of the 
polygon $\mathscr{P}$ admits a normal vector which differs from the 
one of the $(n+1)$-th segment. In other words, the generalized normal 
\eqref{generalized-normal-hyperbolic} defines a discontinuous vector field  
and, in the limit of large $n$, assumes an infinite number of values.  
Furthermore, we note that in our example the reduced boundary $\partial^{*}E$ 
of the polygon $\mathscr{P}$ has a form which agrees with De Giorgi 
structure theorem (cf. Eq. \eqref{(1.1)}).

\section{An example of generalized normal in elliptic geometry} 
\setcounter{equation}{0}
\label{Sec:elliptic-metric}

In Euclidean geometry, given a line and a point which does not lie on this line, 
there exists one and only one line which passes through the given point 
and is parallel to the given line. On the other hand, in hyperbolic geometry, 
that we have considered in the previous section, infinitely many distinct  
parallel lines can be found. A third option, where no parallel lines exist, 
gives rise to elliptic geometry \cite{Coxeter1998} and will be considered here. 
The most common model of elliptic geometry is represented by the surface of a 
sphere (however, it should be noted that in elliptic geometry two lines are usually 
assumed to intersect at a single point, while in spherical geometry two 
great circles intersect at two points, which is why spherical geometry is
said to be a doubly elliptic geometry). Another relevant example of 
(two-dimensional) elliptic geometry is Klein's conformal model of 
the elliptic plane \cite{Coxeter1998}.  
 
In this section, we consider an example of two-dimensional elliptic 
geometry having squared line element \cite{Liebscher2005} 
\begin{align}
{\rm d}s^2 = \dfrac{{\rm d}x^2 + {\rm d}y^2 + \left(x {\rm d}y-y 
{\rm d}x\right)^2}{\left(1+x^2+y^2\right)^2}.
\label{elliptic-metric}
\end{align}
The resulting connection coefficients read as 
\begin{align}
\Gamma^{x}_{\phantom{x}xx}&=-\dfrac{2x}{\left(1+x^2+y^2\right)}, 
\nonumber \\
\Gamma^{x}_{\phantom{x}xy}&=-\dfrac{y}{\left(1+x^2+y^2\right)},
\nonumber \\
\Gamma^{y}_{\phantom{y}xy}&=-\dfrac{x}{\left(1+x^2+y^2\right)},
\nonumber \\
\Gamma^{y}_{\phantom{y}yy}&=-\dfrac{2y}{\left(1+x^2+y^2\right)},
\end{align}
and hence the geodesic equations take the form
\begin{subnumcases}{\label{geod-eqs-elliptic}}
x^{\prime \prime} =   \dfrac{2 x^{\prime}}{\left(1+x^{2}+y^{2}\right)} 
\left(x x^{\prime} + y y^{\prime}\right),
& \label{elliptic-geod-eq-1}\\
y^{\prime \prime} =   \dfrac{2 y^{\prime}}{\left(1+x^{2}+y^{2}\right)} 
\left(x x^{\prime} + y y^{\prime}\right),
& \label{elliptic-geod-eq-2}
\end{subnumcases}
where the prime denotes differentiation with respect to the affine parameter 
$\tau$. If we eliminate the factor $(x x^{\prime} + y y^{\prime})/(1+x^{2}+y^{2})$ 
from the second equation and substitute it in the first, we get
\begin{align}
\dfrac{{\rm d}}{{\rm d}\tau} \left[ \log \left(\dfrac{x^\prime}{y^\prime}\right)\right]=0,
\end{align}
which leads to
\begin{align}
x(\tau)=\alpha_1 y(\tau)+ \alpha_2, \quad (\alpha_1 \neq 0, \alpha_2 \in \mathbb{R}).
\label{x-tau-and-y-tau}
\end{align}
Then, from Eqs. \eqref{elliptic-geod-eq-2} and \eqref{x-tau-and-y-tau} 
we obtain the geodesic solutions
\begin{align}
x(\tau) &=\dfrac{\alpha_2 + \left(\alpha_1 \sqrt{1+\left(\alpha_1\right)^2 
+\left(\alpha_2\right)^2}\right) \tan \left[c_1 \left(\tau+c_2\right)\sqrt{1
+\left(\alpha_1\right)^2 +\left(\alpha_2\right)^2} \right]}{1+\left(\alpha_1\right)^2}, 
\label{solution-x-tau}
\\
y(\tau) &=\dfrac{-\alpha_1\alpha_2 + \left( \sqrt{1+\left(\alpha_1\right)^2 
+\left(\alpha_2\right)^2}\right) \tan \left[c_1 \left(\tau+c_2\right)\sqrt{1
+\left(\alpha_1\right)^2 +\left(\alpha_2\right)^2} \right]}{1+\left(\alpha_1\right)^2}, 
\label{solution-y-tau}
\end{align} 
where $c_1$ and $c_2$ are integration constants. 

We note that the system \eqref{geod-eqs-elliptic} is not affected if we 
interchange the role of $x$ and $y$, i.e., it is invariant under the transformation
\begin{align}
  x \leftrightarrow y.
\end{align}
This means that 
\begin{align}
x_2(\tau) &=\dfrac{-\tilde{\alpha}_1\tilde{\alpha}_2 + \left( \sqrt{1
+\left(\tilde{\alpha}_1\right)^2 +\left(\tilde{\alpha}_2\right)^2}\right) 
\tan \left[\tilde{c}_1 \left(\tau+\tilde{c}_2\right)\sqrt{1+\left(\tilde{\alpha}_1\right)^2 
+\left(\tilde{\alpha}_2\right)^2} \right]}{1+\left(\tilde{\alpha}_1\right)^2}, 
\\
y_2(\tau) &=\dfrac{\tilde{\alpha}_2 + \left(\tilde{\alpha}_1 \sqrt{1
+\left(\tilde{\alpha}_1\right)^2 +\left(\tilde{\alpha}_2\right)^2}\right) 
\tan \left[\tilde{c}_1 \left(\tau+\tilde{c}_2\right)\sqrt{1+\left(\tilde{\alpha}_1\right)^2 
+\left(\tilde{\alpha}_2\right)^2} \right]}{1+\left(\tilde{\alpha}_1\right)^2}, 
\end{align} 
is a solution of Eq. \eqref{geod-eqs-elliptic}, where $\tilde{c}_1,\tilde{c}_2$ 
are real-valued integration constants and 
\begin{align}
y_2(\tau)=\tilde{\alpha}_1 x_2(\tau)+ \tilde{\alpha}_2, \quad (\tilde{\alpha}_1 \neq 0, 
\tilde{\alpha}_2 \in \mathbb{R}).
\end{align}

\begin{figure}
    \centering
    \includegraphics[scale=0.72]{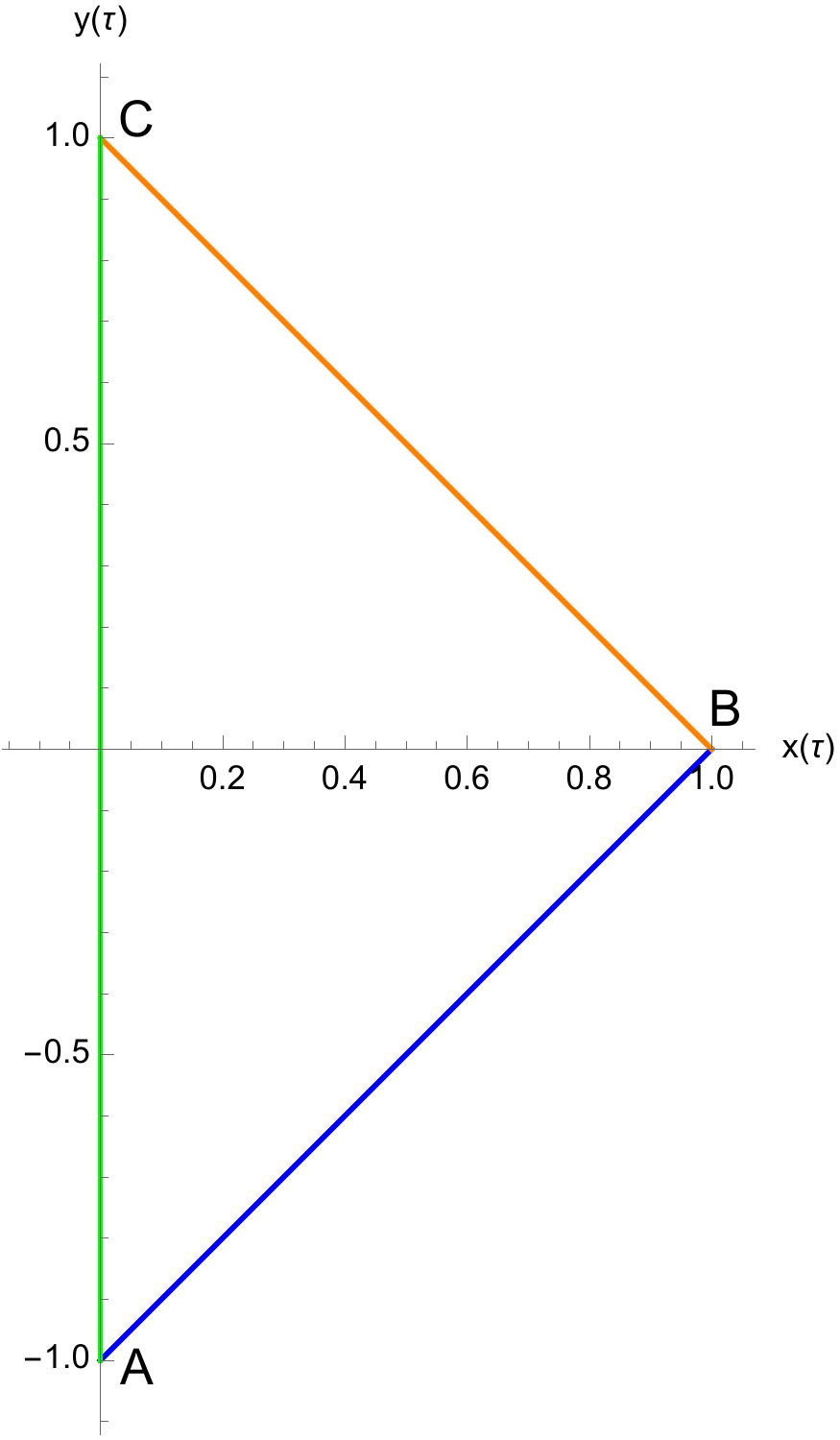}
    \caption{The geodesic triangle $\mathscr{T}$ having vertices 
\eqref{vertices-geod-triangle} and parametrized by Eqs. \eqref{sideAB}, \eqref{sideBC}, and \eqref{sideCA}.}
    \label{fig:geod-triangle}
\end{figure}

We can now set up a geodesic triangle $\mathscr{T}$ by means of Eqs. 
\eqref{x-tau-and-y-tau}--\eqref{solution-y-tau}. Let
\begin{align}
A &\equiv (0,-1),
\nonumber \\
B &\equiv (1,0),
\nonumber \\
C &\equiv (0,1),
\label{vertices-geod-triangle}
\end{align}
be the vertices of $\mathscr{T}$ (see Fig. \ref{fig:geod-triangle}). 
Then, the side $\mathscr{T}_{AB}$ of $\mathscr{T}$ connecting $A$ and $B$ is parametrized by
\begin{align}
\mathscr{T}_{AB}: \left \{ 
\begin{array}{rl}
& x_{AB}(\tau)=\dfrac{1}{2}\left\{1+\sqrt{3} \tan \left[2 \sqrt{3}\left(3+\tau\right)\right]\right\},
\\[0.3cm]
& y_{AB}(\tau)=\dfrac{1}{2}\left\{-1+\sqrt{3} \tan \left[2 \sqrt{3}\left(3+\tau\right)\right]\right\},
\\[0.3cm]
& \tau_1 \leq \tau \leq \tau_2,
\end{array}
\right.
\label{sideAB}
\end{align}
where 
\begin{align}
\tau_1 &\equiv -\dfrac{\left(6 \sqrt{3} + \pi/6\right)}{2\sqrt{3}},
\nonumber \\
\tau_2 &\equiv -\dfrac{\left(6 \sqrt{3} - \pi/6\right)}{2\sqrt{3}},
\end{align}
and the functions $ x_{AB}(\tau)$ and $ y_{AB}(\tau)$ can be obtained from Eqs. 
\eqref{solution-x-tau} and \eqref{solution-y-tau}, respectively, by setting  
\begin{subequations}
\begin{align}
\alpha_1 &=\alpha_2=1,
\label{side-AB-alpha1-2}
\\
c_1&=2, 
\\
c_2&=3.
\end{align}
\end{subequations}
The side $\mathscr{T}_{BC}$, which links the vertices $B$ and $C$, is defined by
\begin{align}
\mathscr{T}_{BC}: \left \{ 
\begin{array}{rl}
& x_{BC}(\tau)=\dfrac{1}{2}\left\{1-\sqrt{3} \tan \left[2 \sqrt{3}\left(3+\tau\right)\right]\right\},
\\[0.3cm]
& y_{BC}(\tau)=\dfrac{1}{2}\left\{1+\sqrt{3} \tan \left[2 \sqrt{3}\left(3+\tau\right)\right]\right\},
\\[0.3cm]
& \tau_1 \leq \tau \leq \tau_2.
\end{array}
\right.
\label{sideBC}
\end{align}
The functions $ x_{BC}(\tau)$ and $ y_{BC}(\tau)$ can be read off from Eqs. 
\eqref{solution-x-tau} and \eqref{solution-y-tau}, respectively, with 
\begin{subequations}
\begin{align}
\alpha_1 &=-\alpha_2=-1,
\label{side-BC-alpha1-2}
\\
c_1&=2, 
\\
c_2&=3.
\end{align}
\end{subequations}
Last, $\mathscr{T}_{CA}$ can be obtained from
\begin{align}
\mathscr{T}_{CA}: \left \{ 
\begin{array}{rl}
& x_{CA}(\tau)=0,
\\[0.3cm]
& y_{CA}(\tau)=-\tan \left[2 \left(-3+\tau\right)\right],
\\[0.3cm]
& \tau_3 \leq \tau \leq \tau_4,
\end{array}
\right.
\label{sideCA}
\end{align}
with
\begin{align}
\tau_3 & \equiv \dfrac{1}{2} \left(6-\pi/4\right),
\nonumber \\
\tau_4 & \equiv \dfrac{1}{2} \left(6+\pi/4\right),
\end{align}
the functions $x_{CA}(\tau)$ and $y_{CA}(\tau)$ stemming Eqs. \eqref{solution-x-tau} 
and \eqref{solution-y-tau}, respectively, when 
\begin{subequations}
\begin{align}
\alpha_1 &=\alpha_2=0,
\\
c_1&=-2, 
\\
c_2&=-3.
\end{align}
\end{subequations}
It should be noted that the solution having $\alpha_1=0$ can be regarded  
as a \qm{limiting} case of Eqs. \eqref{x-tau-and-y-tau}--\eqref{solution-y-tau}.

At this stage, we are ready to evaluate the normal vector field to the geodesic 
triangle $\mathscr{T}$. Given the curve $y=f(x)$, it is known that the 
equation of its normal to a generic point having coordinates $(X,Y)$ is \cite{Tortotici}
\begin{equation}
y-Y=-\dfrac{1}{f^{\prime}(X)} \left(x-X\right),
\label{normal-equation-Tortotici}
\end{equation}
the prime denoting differentiation with respect to the $x$ variable. In our 
example, the geodesics can be equivalently described by means of (cf. Eq. \eqref{x-tau-and-y-tau})
\begin{align}
y(x)=\dfrac{x}{\alpha_1}-\dfrac{\alpha_2}{\alpha_1}, \qquad \quad (\alpha_1 
\neq 0, \alpha_2 \in \mathbb{R}),
\label{y-f-of-x-equation}
\end{align}
which means that, recalling Eq. \eqref{side-AB-alpha1-2}, the equation of 
the side $\mathscr{T}_{AB}$ reads also as\footnote{It is clear that Eq. 
\eqref{side-AB-y-of-x} can be obtained also from  Eq. \eqref{sideAB}.}
\begin{align}
\mathscr{T}_{AB}: \quad y_{ AB} \left(x\right)= x-1, \qquad x \in [0,1],
\label{side-AB-y-of-x}
\end{align}
and hence by virtue of Eq. \eqref{normal-equation-Tortotici} the normal 
vector field $\mathscr{N}_{AB}$  to $\mathscr{T}_{AB}$ is defined by
\begin{align}
\mathscr{N}_{AB}: \quad y_{n_{AB}}\left(x\right) =-x + \bar{x}+ \bar{y},
\label{normal-to-AB}
\end{align}
where $(\bar{x},\bar{y})$ are the coordinates of a generic point belonging 
to $\mathscr{T}_{AB}$.  As a consequence of  \eqref{side-BC-alpha1-2} 
(or, equivalently, Eq. \eqref{sideBC}), we have for the side $\mathscr{T}_{BC}$
\begin{align}
\mathscr{T}_{BC}: \quad y_{BC} \left(x\right)= -x+1, \qquad x \in [0,1],
\end{align}
from which we derive the form of its normal vector field $\mathscr{N}_{BC}$
\begin{align}
\mathscr{N}_{BC}: \quad y_{n_{BC}}\left(x\right) =x - \hat{x}+ \hat{y},
\label{normal-to-BC}
\end{align}
$(\hat{x},\hat{y})$ being the coordinates of a point lying on $\mathscr{T}_{BC}$.  
As pointed out before, the side $\mathscr{T}_{CA}$ cannot be obtained directly 
from \eqref{y-f-of-x-equation}. Indeed, its equation is 
\begin{align}
\mathscr{T}_{CA}: \left \{ 
\begin{array}{rl}
& x_{CA}=0,
\\[0.3cm]
& -1 \leq y\leq 1,
\end{array}
\right.
\end{align}
which means that $\mathscr{N}_{CA}$ is
\begin{align}
\mathscr{N}_{CA}: \quad y_{n_{CA}} =k, \qquad -1 \leq k \leq 1.
\label{normal-to-CA}
\end{align}

Likewise the hyperbolic example discussed in Sec. \ref{Sec:hyperbolic-plane},  
Eqs. \eqref{normal-to-AB}, \eqref{normal-to-BC}, and \eqref{normal-to-CA} 
give rise to a discontinuous normal vector field.

\section{Application to two-dimensional Euclidean quantum gravity} 
\setcounter{equation}{0}
\label{Sec:Application}

As far as we know, the kind of thinking used so far in our paper was 
lacking in the literature on fundamental interactions in physics.
In recent times, much work in the mathematical literature 
has been devoted to the investigation
of geometric measure theory in non-Euclidean spaces \cite{SNS}.
Within such a framework, the occurrence of discontinuous normal
vector fields has to be considered \cite{book5}, and our original
explicit examples in Secs. \ref{Sec:hyperbolic-plane} and 
\ref{Sec:elliptic-metric} can be of interest. 

On the other hand, the examples of Secs. \ref{Sec:hyperbolic-plane} and 
\ref{Sec:elliptic-metric} have a clear mathematical motivation, but they
do not have an impact on theoretical physics. 
For this purpose, we are currently considering
the case of Euclidean quantum gravity 
\cite{1993,1994,1996,1997,1998,1999,2005a,2005b,2021a,2021b}. 
Within this framework, a prescription
for functional integration is necessary, and we propose to consider only 
finite-perimeter sets that match the assigned data on their reduced boundary. 
We arrive at this prescription upon bearing in mind that measurable sets
belong to two families: either they have finite perimeter, or they do not.
The restriction to finite-perimeter sets might be severe, but it leads
to mathematical properties which are under control and hence merits a
careful assessment.

A second choice is also in order, and it has to do with the dimension of
such finite-perimeter sets. In higher dimensions a problem arises, i.e., how to 
define a vector that plays the role of normal. Even just for a straight line
in three-dimensional Euclidean space, what is defined is the plane orthogonal
to such a line, but there is no coordinate-independent way of selecting two
linearly independent normal vector fields to the line. Thus, a four-dimensional
analogue of our Secs. \ref{Sec:hyperbolic-plane} and \ref{Sec:elliptic-metric}  
is not conceivable, as far as we can see. 
We have therefore resorted to two-dimensional Euclidean quantum gravity,  
by focusing on the Euclidean two-dimensional  
CGHS dilaton gravity model \cite{CGHS1992,Harvey1992,Strominger1995} (see also e.g. Ref. 
\cite{Bergamin2005} for a general framework pertaining to Euclidean dilaton gravity in 
two dimensions). This example will clarify
the role of the discontinuous normal $\nu_E$ occurring in the divergence theorem \eqref{(1.14)}.

The Euclidean CGHS action reads as
\begin{align} \label{CGHS-action}
\mathcal{I}_E=-\dfrac{1}{2\pi} \int \d^2 x \sqrt{g} \, {\rm e}^{-2 \phi} \left( R 
+ 4 \nabla^\mu \phi \nabla_\mu \phi + 4 \Lambda^2 \right),
\end{align}
$g$ denoting the metric determinant, $R$  the Ricci scalar, $\phi$ the dilaton field, and 
$\Lambda^2$ a cosmological constant. Bearing in mind the recipes of Ref. \cite{Lee1990} 
(see also Refs. \cite{Jacobson1994,Myers1994}),  we can write the variations of the Lagrangian 
$\mathcal{L}_E$ occurring in Eq. \eqref{CGHS-action} as
\begin{align}
\delta \left(\sqrt{g}\,\mathcal{L}_E\right)
=- \dfrac{\sqrt{g}}{2 \pi} \left[\left(E_g\right)_{\mu \nu} 
\delta g^{\mu \nu} + E_\phi \delta \phi + \nabla_{\alpha} \Theta^{\alpha}\right],
\end{align}
where
\begin{align}
\left(E_g\right)_{\mu \nu} &=0, 
\nn \\
E_\phi &=0,
\end{align}
are the metric and the dilaton field equations, respectively, and  $\Theta^\alpha$ is referred 
to as the  symplectic potential current density. By virtue of Eq. \eqref{CGHS-action}, we find
\begin{subequations}
\begin{align}
\left(E_g\right)_{\mu \nu} &= 2 {\rm e}^{-2 \phi} \left[ g_{\mu \nu} \left((\nabla_\alpha \phi) 
(\nabla^\alpha \phi) - \Lambda^2 -\nabla_\alpha \nabla^\alpha \phi \right) 
+ \nabla_\mu \nabla_\nu \phi \right],
\\
E_\phi &=-2 {\rm e}^{-2 \phi} \left( R + 4 \nabla_\alpha \nabla^\alpha \phi + 4 \Lambda^2 
-4 \nabla_\alpha \phi \nabla^\alpha \phi\right),
\\
\Theta^\alpha &=  {\rm e}^{-2 \phi} \bigl[ 8 \left(\nabla^\alpha \phi\right) \delta \phi 
- \nabla_\mu \delta g^{\mu \alpha} + g_{\mu \nu} \nabla^\alpha \delta g^{\mu \nu} 
-2 \left(\nabla_\mu \phi \right) \delta g^{\mu \alpha} 
\nn \\
&+ 2 \left(\nabla^\alpha \phi \right) g_{\mu \nu} \delta g^{\mu \nu} \bigr].
\end{align}
\end{subequations}

We aim at evaluating the integral
\begin{align} \label{integral-of-nabla-theta}
-\dfrac{1}{2 \pi} \int_\mathcal{C} \d^2 x \sqrt{g} \, \nabla_\alpha \Theta^\alpha,
\end{align}
where $\mathcal{C}$ is a two-dimensional region having finite perimeter. 
For this purpose, we find it more convenient to consider an off-shell calculation,
i.e. for metric and dilaton field which do not obey the Euclidean field 
equations. In such a way, we may choose an example 
as close as possible to the ones developed in the theory of
finite-perimeter sets in $n$-dimensional Euclidean space \cite{book6}, while avoiding
the difficult task of solving the coupled partial differential equations 
for metric and dilaton. 

\subsection{Preparing the ground for evaluating the integral \eqref{integral-of-nabla-theta}}

In order to prepare the ground for investigating the integral \eqref{integral-of-nabla-theta},
let us consider the following example of two-dimensional finite-perimeter set \cite{book6}, which provides  
a slight modification of the case discussed in Sec. \ref{Sec:Intro} (which we recall is valid in 
$\mathbb{R}^n$ with $n \geq 3$, see Eq. \eqref{(1.12)}). Consider in ${\mathbb{R}}^{2}$ the set $E$ defined by
\begin{align}
E =\cup_{k =1}^{\infty} \, B_{r_k}(x_k)
\end{align}
where, as before, $B_{r_k}(x_k)$ denotes the Euclidean open ball centred at the point 
$x_k \in \mathbb{R}^2$ and having small positive radius $r_k$, which is supposed to  satisfy
\begin{align}
  r_k < \varepsilon,  \;\; \forall\,k \in \mathbb{N},
\end{align}
such that 
\begin{align}
2 \omega_2 \sum \limits_{k=1}^{\infty} r_k \leq 1,
\label{Maggi-inequality}
\end{align}
with 
\begin{align}
\omega_2 = \dfrac{\Gamma(1/2)^2}{\Gamma(1)}= \pi,
\end{align}
$\Gamma(n)$ being the Euler $\Gamma$-function. The perimeter of $B_{r_k}(x_k)$ is
\begin{align}
P\biggl(B_{r_k}(x_k)\biggr)=H^1\biggl(\partial B_{r_k}(x_k)\biggr) = 2 \omega_2 r_k,
\label{Maggi-1}
\end{align}
$H^1$ denoting the one-dimensional Hausdorff  measure.  Starting from Eq. \eqref{Maggi-1}, it can be shown 
\cite{book6} that for every $N \in \mathbb{N}$, the set
\begin{align}
E_N = \sum_{k=1}^{N} B_{r_k}(x_k),
\end{align}
has finite perimeter since, by virtue of Eq. \eqref{Maggi-inequality}, 
\begin{align}
P(E_N) \leq \sum_{k=1}^{N} P\biggl(B_{r_k}(x_k)\biggr) \leq 2 \omega_2 \sum_{k =1}^{\infty} r_k \leq 1.
\label{Maggi-2}
\end{align}
Therefore, as $\vert E \vert \leq \omega_2 < \infty$ ($\vert E \vert$ being the volume of $E$), 
we have $E_N \to E $ as $N \to \infty$ and hence
\begin{align}
P(E) \leq 1,
\end{align} 
which means that $E$ has finite perimeter. Furthermore, it is possible to  prove that $\vert \partial E 
\vert \geq  \omega_2 - \varepsilon$, which implies as a consequence $H^{1}(\partial E) = \infty$ \cite{book6}.

We can now extend the previous example to a two-dimensional analytic Riemannian manifold $\mathscr{M}$. 
For this reason, let us consider the two-dimensional geodesic ball $B^G_r(p)$ having small positive 
radius $r$ and centred at the point $p \in \mathscr{M}$. The volume $V_r(p)$ of $B^G_r(p)$ 
can be written by means of the power series \cite{Gray}
\begin{align}
V_r(p) &= \omega_2 r^2 \Biggl[ 1-\dfrac{R}{24}r^2 + \dfrac{r^4}{8640} \left(-3 R^{\mu \nu \rho \sigma} 
R_{\mu \nu \rho \sigma} + 8 R^{\mu \nu} R_{\mu \nu} + 5 R^2 -18 \nabla_\mu \nabla^\mu R \right)
\nn \\
&+ {\rm O}\left(r^6\right) \Biggr]_p,
\label{Grey-formula-1}
\end{align}
whereas the volume $S_r(p)$ of $\partial B^G_r(p)$ reads as \cite{Gray}
\begin{align}
S_r(p) &= 2\omega_2 r \Biggl[ 1-\dfrac{R}{12}r^2 + \dfrac{r^4}{2880} \left(-3 R^{\mu \nu \rho \sigma} 
R_{\mu \nu \rho \sigma} + 8 R^{\mu \nu} R_{\mu \nu} + 5 R^2 -18 \nabla_\mu \nabla^\mu R \right)
\nn \\
&+ {\rm O}\left(r^6\right) \Biggr]_p.
\label{Grey-formula-2}
\end{align}
If the trace $R$ of the Ricci tensor is positive, then it follows from Eqs. \eqref{Grey-formula-1} 
and \eqref{Grey-formula-2} that, for sufficiently small $r$,
\begin{align}
V_r(p)  < \omega_2 r^2,
\end{align}
and 
\begin{align}
S_r(p)  < 2\omega_2 r,
\label{Gray-inequality-2}
\end{align}
respectively. 

At this stage, we have  the necessary ingredients to generalize the previous example to the case 
when the manifold $\mathscr{M}$ admits a positive trace $R$ of Ricci. Indeed, along the same lines as before, let 
\begin{align}
\mathcal{C} =\cup_{k =1}^{\infty} \, B^G_{r_k}(x_k),
\end{align}
be a surface of $\mathscr{M}$ constructed in terms of the geodesic balls $B^G_{r_k}(x_k)$ whose radius 
$r_k$ is subject to the condition \eqref{Maggi-inequality}. Bearing in mind Eq. \eqref{Gray-inequality-2}, we have
\begin{align}
P\biggl(B^G_{r_k}(x_k)\biggr)=H^1\biggl(\partial B^G_{r_k}(x_k)\biggr) < 2 \omega_2 r_k,
\end{align}
which makes it possible for us to prove that the set
\begin{align}
\mathcal{C}_N = \sum_{k=1}^{N} B^G_{r_k}(x_k),
\end{align}
has finite perimeter, since, similarly to Eq. \eqref{Maggi-2},
\begin{align}
P(\mathcal{C}_N) \leq \sum_{k=1}^{N} P\biggl(B_{r_k}(x_k)\biggr) < 2 \omega_2 \sum_{k =1}^{\infty} r_k \leq 1.
\end{align}
Therefore, the set $\mathcal{C}$, which can be obtained from $\mathcal{C}_N$ in the limit $N \to \infty$, has finite perimeter with 
\begin{align}
P(\mathcal{C}) <1,
\end{align}
although $H^{1}(\partial \mathcal{C}) = \infty$.

\section{Concluding remarks}
\label{Sec:Conclusions}

By means of the methods presented in our paper, 
it is possible to evaluate the integral 
\eqref{integral-of-nabla-theta} over the finite-perimeter set $\mathcal{C}$,
and we hope to have outlined the good potentialities of geometric measure
theory for a fresh look at the unsolved problems of traditional formulations
of Euclidean quantum gravity. 

The next task wil be the rigorous proof of existence theorems of
Euclidean functional integrals for gravity in two and higher dimensions,
when restricted to finite-perimeter sets. If it were possible to accomplish
this, the associated reduced boundary construction might open a new era
in quantum gravity and quantum cosmology.

\section*{acknowledgments}
E. B. is  grateful to Nicola Fusco for invaluable discussions regarding many relevant 
topics of geometric measure theory. The authors thank Marco Abate for
enlightening correspondence. This work is supported by the Austrian Science Fund 
(FWF) grant P32086. G. E. dedicates this research to Margherita.

\begin{appendix}

\section{Fi\-ni\-te-pe\-ri\-me\-ter se\-ts a\-nd the\-ir re\-du\-ced bo\-un\-da\-ry}
\setcounter{equation}{0}
\label{Appendix:Finite-Perimeter-Set}

On denoting by $\varphi(x,E)$ the characteristic function of
a set $E \subset {\mathbb R}^{n}$ (by definition, $\varphi(x,E)$ equals $1$
at points $x \in E$, and $0$ otherwise) and by $*$
the convolution product of two functions defined on ${\mathbb{R}}^{n}$:
\begin{equation}
f*h(x) \equiv \int f(x-\xi) h(\xi) \, \d\xi,
\label{(A.1)}
\end{equation}
De Giorgi defined \cite{DG1,DG2,book2} for all integer $n \geq 2$ and for all
$\lambda >0$ the function
\begin{equation}
\varphi_{\lambda}: x \rightarrow 
\varphi_{\lambda}(x) \equiv (\pi \lambda)^{-{n \over 2}}
{\rm exp} \left(-{\sum\limits_{k=1}^{n}(x_{k})^{2} \over \lambda}\right)
* \varphi(x,E),
\label{(A.2)}
\end{equation}
and, as a next step, the {\it perimeter} of the set
$E \subset {\mathbb R}^{n}$
\begin{equation}
P(E) \equiv \lim_{\lambda \to 0} \int_{{{\mathbb R}}^{n}}
\sqrt{\sum_{k=1}^{n} \left({\partial \varphi_{\lambda}
\over \partial x_{k}}\right)^{2}} \; \d x.
\label{(A.3)}
\end{equation}
The perimeter defined in Eq. \eqref{(A.3)} is not always finite. A
necessary and sufficient condition for $P(E)$ to be finite is
the existence of a set function of vector nature completely 
additive and bounded, defined for any set $B \subset {\mathbb R}^{n}$
and denoted by $a(B)$, verifying the generalized Gauss-Green formula
\begin{equation}
\int_{E}Dh \; \d x=-\int_{{\mathbb R}^{n}}h(x) \; \d a.
\label{(A.4)}
\end{equation}
If Eq. \eqref{(A.4)} holds, the function $a$ is said to be the distributional
gradient of the characteristic function $\varphi(x,E)$. 
A {\it polygonal domain} is every set $E \subset {\mathbb R}^{n}$ that
is the closure of an open set and whose topological boundary 
$\partial E$ is contained in the union of a finite number of
hyperplanes of ${\mathbb R}^{n}$. The sets approximated by polygonal
domains having finite perimeter were introduced by Caccioppoli
\cite{RC1,RC2} and coincide with the collection of all 
finite-perimeter sets \cite{book2}. This is why finite-perimeter 
sets are said to be Caccioppoli sets. 

Is $E$ is a finite-perimeter set of ${\mathbb R}^{n}$, its reduced boundary
$\partial^{*}E$ is the collection of all points $\xi$ for which:
\vskip 0.3cm
\noindent
(i) The integral of the norm of the gradient of the characteristic function,
when taken over the ball (our ball is called open hypersphere by De Giorgi) 
centred at $\xi$ and of radius $\rho$, is always positive, i.e.
\begin{equation}
\int_{B_{\rho}(\xi)} \left \| {\rm grad} \; \varphi(x,E) \right \| > 0 \;
\forall \rho>0,
\label{(A.5)}
\end{equation}
(ii) The limit defining the normal vector exists, i.e.
\begin{equation}
\lim_{\rho \to 0}{\int_{B_{\rho}(\xi)}{\rm grad} \varphi(x,E) \over
\int_{B_{\rho}(\xi)}\left \| {\rm grad} \varphi(x,E) \right \|}=\nu(\xi),
\label{(A.6)}
\end{equation}
(iii) The resulting normal vector has unit Euclidean norm.

\section{Calculation of the hyperbolic distance 
in the limit of large $n$} 
\setcounter{equation}{0}
\label{Appendix-log-2}

In Sec. \ref{Sec:hyperbolic-plane}, we have evaluated the large-$n$ limit of the 
hyperbolic distance  $\rho_{\mathbb{H}} \left(z_{n-1},z_{n}\right)$ between the 
vertices $z_{n-1}$ and $z_{n}$ of the $(n+2)$-sided  hyperbolic polygon $\mathscr{P}$, 
defined by Eq.  (\ref{points-hyperbolic-polygon}),  by employing Eq. (\ref{hyperbolic_distance-2}). 
The final result of this computation is displayed in  Eq. 
(\ref{distance-z-n-1-and-z-n-large-n}). In this appendix, we show how the 
application of Eq. (\ref{hyperbolic_distance-1}) for this calculation would have led to a misleading result. 

Bearing in mind Eqs. (\ref{point-z-n-1}) and (\ref{point-z-n}), from 
Eq. (\ref{hyperbolic_distance-1}) we obtain 
\begin{equation}
\rho_{\mathbb{H}} \left(z_{n-1},z_{n}\right) = \log \left( \dfrac{1+\sqrt{\dfrac{2^{2n}
+3^{2n}}{2^{2n}+3^{2n+2}}}}{1-\sqrt{\dfrac{2^{2n}+3^{2n}}{2^{2n}+3^{2n+2}}}} \right),
\end{equation}
and hence, in the limit of large $n$, the above equation yields
\begin{equation}\label{distance-log-2}
\lim_{n \to \infty} \rho_{\mathbb{H}} \left(z_{n-1},z_{n}\right)  
= \log \left(\dfrac{1+\sqrt{\dfrac{1}{9}}}{1-\sqrt{\dfrac{1}{9}}}\right) = \log 2.
\end{equation}
Therefore, in Eq. (\ref{distance-log-2}) we recover a different result from the one given  
in Eq. (\ref{distance-z-n-1-and-z-n-large-n}). As pointed out before, this is due to the 
fact that the vertices $z_{n-1}$ and $z_{n}$ of the polygon $\mathscr{P}$ do not belong, 
in the limit of large $n$,  to the hyperbolic plane $\mathbb{H}$, and hence  Eq. 
(\ref{hyperbolic_distance-1}) becomes meaningless under this hypothesis. 

\end{appendix}

\end{document}